# Efficient Reachability Query Evaluation in Large Spatiotemporal Contact Datasets

Houtan Shirani-Mehr
Computer Science Dept.
Univ. of Southern California
hshirani@usc.edu

Farnoush Banaei-Kashani
Computer Science Dept.
Univ. of Southern California
banaeika@usc.edu

Cyrus Shahabi
Computer Science Dept.
Univ. of Southern California
shahabi@usc.edu

## ABSTRACT

With the advent of reliable positioning technologies and prevalence of location-based services, it is now feasible to accurately study the propagation of items such as infectious viruses, sensitive information pieces, and malwares through a population of moving objects, e.g., individuals, mobile devices, and vehicles. In such application scenarios, an item passes between two objects when the objects are sufficiently close (i.e., when they are, so-called, *in contact*), and hence once an item is initiated, it can penetrate the object population through the evolving network of contacts among objects, termed *contact network*. In this paper, for the first time we define and study reachability queries in large (i.e., disk-resident) contact datasets which record the movement of a (potentially large) set of objects moving in a spatial environment over an extended time period. A reachability query verifies whether two objects are "reachable" through the evolving contact network represented by such contact datasets. We propose two contact-dataset indexes that enable efficient evaluation of such queries despite the potentially humongous size of the contact datasets. With the first index, termed *ReachGrid*, at the query time only a small necessary portion of the contact network which is required for reachability evaluation is constructed and traversed. With the second approach, termed *ReachGraph*, we precompute reachability at different scales and leverage these precalculations at the query time for efficient query processing. We optimize the placement of both indexes on disk to enable efficient index traversal during query processing. We study the pros and cons of our proposed approaches by performing extensive experiments with both real and synthetic data. Based on our experimental results, our proposed approaches outperform existing reachability query processing techniques in contact networks by 76% on average.

## 1. INTRODUCTION

Studying how items such as infectious viruses, ideas and habits, malwares, and broadcast messages propagate through a population of moving objects, e.g., individuals, mobile devices, or vehicles, is of importance in a wide range of applications including public health monitoring, social behavior analysis, computer security and intelligent traffic monitoring, to name a few. In such application scenarios, objects pass items among themselves once they are in sufficiently close distance, i.e., once they are so called in *contact*. Accordingly, once an item is initiated by an object, it can penetrate the evolving network of contacts among objects termed the *contact network*. With such analysis, one can for instance, design public health interventions in order to control propagation of infectious diseases, or find the source(s) that have originally leaked sensitive information or initiated spread of malwares.

Arguably, one of the main building blocks for item propagation analysis in evolving contact networks is the ability to compute reachability queries which evaluate whether two objects are "reachable" through the evolving contact network. Previously, lack of accurate datasets that capture the contact networks has limited the accuracy and applicability of propagation analysis (and particularly, reachability analysis) in contact networks, and previous studies have inevitably resorted to simplified contact network models, or small-scale and inaccurate contact datasets. However, with the recent advances in developing accurate positioning devices and prevalence of location-based services, it is becoming possible to capture the location of objects in large scales and for extended periods of time, resulting in very large contact datasets that capture the history of objects contacts accurately and with high spatiotemporal resolution. In this paper, we focus on defining and efficient evaluation of reachability queries in large-scale (disk-resident) historic contact datasets, where the main challenge is to reduce the computation time for query evaluation.

Consider the contact network depicted in Figure 1 which shows the position of a set of objects at each time instance within the time interval $T=[0,3]$. In this figure, two objects are connected by a link if they are in contact; for instance, $o_1$ and $o_2$ are in contact at time 0. The object $o_4$ is reachable from $o_1$ during time interval of $[0, 1]$. The reason is that if an item initiated by $o_1$ at time 0, it can pass from $o_1$ to $o_2$ at time 0 and then from $o_2$ to $o_4$ at time 1. Note that in the same figure, $o_1$ is not reachable from $o_4$ during $[0, 1]$. Consider the following examples on how reachability query evaluation plays a fundamental role in analyzing item propagation through contact networks in the context of some application scenarios mentioned above. With the first example, assume a set of individuals $O$ are known to carry a dangerous contagious virus. By performing a batch of reachability queries between each individual in $O$ and the rest of the population, the individuals who could have been directly or indirectly contaminated within a certain time interval can be identified by determining the set of individuals reachable from $O$ in the same time interval. Note that this application requires running potentially numerous reachability queries between pairs of individuals which can be very time consuming. On the other hand, timely medication administration can save lives with most viral dis-eases. Next, imagine a set of individuals $O$, e.g., criminals, on a watch list and need to be monitored. Law enforcement agencies may need to discover those who have been potentially in contact with any of the individuals in $O$. Again, this requires performing batch of reachability queries to find all the individuals reachable from/to any individual in $O$. Such analysis may help in preventing new crimes and to analyze the relationship between criminals.

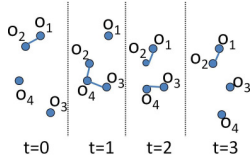

**Figure 1: Objects positions and contacts between them during the time interval [0,3]**

Graph reachability problems which verify whether a path exists between two given vertices of a graph are extensively studied in the recent years [18, 19]. Our problem is different from the existing work on graph reachability in two ways. First, while previous work on graph reachability assumes the graph is memory-resident, we focus on very large disk-resident contact networks. Accordingly, we study how to index the contact network on disk to enable efficient query processing. Second, with our problem objects are associated with time and space information as they move in an environment over time. We show that we can leverage such information for efficient reachability query processing, whereas the existing work on graph reachability only focuses on datasets that are modeled by abstract graphs with no connection to space and time.

In this paper, we propose two index structures for indexing contact networks, namely *ReachGrid* and *ReachGraph*. Consider a reachability query which verifies whether an object (query source) can reach another object (query destination) through the contact network, if we consider only the contacts occurring during a given query time interval (query interval). With ReachGrid, our approach is to compute reachability on-the-fly by expanding the contact network starting from the query source. However, the naïve expansion of the network is prohibitively costly. Instead, to enable "guided" expansion, we leverage the following simple and powerful observation about contact networks; only contacts that occur in the same spatial and temporal locality are relevant for exploration and therefore, exploration of the contacts can be guided through relevant spatiotemporal localities and can avoid other localities for enhanced performance. In particular, with ReachGrid we propose a spatiotemporal grid to index all contacts in the contact network dataset into distinct spatiotemporal localities. At the query time, this index is used to guide on-the-fly expansion of the contact network to verify reachability.

On the other hand, with ReachGraph we use the alternative approach of precomputing the reachability between objects. It is impractical to precompute reachability for all combinations of query source, destination and interval. Therefore, we propose to precompute reachability query only for carefully selected combinations of query source, destination and interval, and leverage these combinations to compute reachability for all other combinations on-the-fly. In turn, at the query time this allows recursively breaking the given reachability query to a set of precomputed reachability queries for efficient query processing.

Finally, with both ReachGrid and ReachGraph, the placement of index on disk can significantly affect the efficiency of query processing. A naïve approach of placing indexes (graph nodes and grid cells) on random disk blocks significantly deteriorate query efficiency. Accordingly guided by the two following observation, we develop enhanced disk placement approaches for ReachGrid and ReachGraph. First, contacts are processed ordered by occurrence time during query processing. Second, during index traversal, an object $o'$ is traversed after $o$, if $o'$ is reachable from $o$. We present our proposed disk placement approaches for ReachGrid and ReachGraph on disk in Sections 4 and 5, respectively.

While ReachGrid evaluates reachability by sweeping contacts along space and time dimensions, ReachGraph computes reachability by traversing a connectivity graph. Accordingly, one can expect ReachGrid to be comparable with ReachGraph when query time interval is small, and vice versa. This expectation is confirmed by our empirical study in Section 6. Moreover, our proposed approaches outperform the existing reachability query processing algorithms by 76% on average.

The rest of the paper is organized as follows. The related work is outlined in Section 2. We formally define reachability query in contact networks in Section 3. We present ReachGrid and ReachGraph indexing techniques in Sections 4 and 5, respectively. Section 6 presents our experimental results. We discuss extensions of our reachability problem in Section 7. Finally, we conclude the paper and discuss the possible future work in Section 8.

## 2. RELATED WORK

We review the related work in four categories: graph reachability, trajectory indexing and trajectory join, external graph traversal and graph indexing and finally, contact networks analysis.

### 2.1 Graph Reachability

Given two vertices $u$ and $v$ in a directed graph $G$, graph reachability verifies whether there is a path from $u$ to $v$ [19, 18]. Although we also reduce our problem to graph reachability by converting the contact network into a hypergraph, our problem is different from previous work on graph reachability in several ways. First, in contrast with the previous work where the focus is on memory-resident graphs, we consider disk-resident graphs. Second, we focus on "spatiotemporal" graphs and accordingly leverage the spatial and temporal properties of such graphs for enhanced index construction and graph traversal. In particular, our graph vertex may represent multiple objects and moreover an object can be associated with multiple vertices. Finally, our proposed multi-resolution graph indexing and bidirectional graph traversal approaches are unique and novel, allowing for unprecedented improvement in the efficiency of state of the art reachability query processing approaches.

### 2.2 Trajectory Join and Trajectory Indexing

The research on moving objects data management has traditionally focused mainly on range and nearest neighbor queries. Recently, trajectory join has also been studied [2, 1]. The problem of Closest-Point-of-Approach (CPA) is proposed and studied in [1]. Given a set of trajectories, CPA finds the pair of objects whose closest distance is less than $d$. Although CPA problem is different from trajectory join, the solution to CPA problem can be adopted to solve trajectory join. Although we use trajectory join algorithms in constructing the contact network, our focus is on indexing a contact network for efficient reachability query processing. Another relevant body of related work on trajectory processing is trajectory indexing [5] which focuses on indexing trajectories for efficient processing of range queries and its variations. In contrast, our problem is how to index a contact network for efficient reachability query processing which is much more complex as compared to range query and its variations.

Shortest path on graphs [13, 7] is another body of related work. Given a graph $G=(V,E)$, the shortest path finds the optimally shortest path assuming a traveling cost between each pair of graph vertices. In contrast, with reachability query we are only interested in verifying whether any contact path exists between two objects.

## 2.3 External Graph Traversal and Graph Indexing

With external memory graph traversal [12, 17], researches have extended the classic graph traversal approaches such as Depth-First-Search (DFS) and Breadth-First-Search (BFS). As mentioned earlier, both DFS and BFS can be leveraged to answer reachability queries. However, with our work we try to avoid unnecessary expansion of the graph nodes by designing an efficient multi-resolution index structure and traversal approaches.

Another category of work focuses on indexing temporal graphs. Time expanded network (TEN) and Time aggregated network (TAN) [14] are two models to represent time varying networks. TEN represents the time dependence by instantiating a snapshot of the network at every time instance. TAN extends TEN where the time varying attributes are further aggregated over edges and vertices. We utilize TEN to initially model a contact network but afterward convert it to a more complex index structure as discussed in Section 5. Recently, [8] studied efficient indexing of spatiotemporal networks represented by TEN. However, in this paper the focus is on indexing techniques to enable efficient processing of route evaluation and retrieval queries as opposed to our work which focuses on the complex reachability query processing.

## 2.4 Contact Networks Analysis

Recent studies [16, 10] have focused on analyzing characteristics of the contact networks such as average contact path length between two objects, or time duration until two objects contact each other again are studied recently. This area of work is orthogonal to our work as we are focusing on indexing a contact network for efficient reachability query processing.

Routing in delay-tolerant networks (DTN) which lack continuous network connectivity is another body of relevant work [9]. The difference between this body of work and our work is two fold. First, the goal of routing in DTN is to find a best path from a source to a destination node based on a cost metric such as messages delivery ratio. Next, our reachability query is associated with a time interval parameter which is leveraged during index construction and query processing to enable efficient reachability query processing.

## 3. PROBLEM DEFINITION

In this section, we first define contact network and afterward formalize the reachability query in a contact network.

### 3.1 Contact Network

Consider a set of objects $O$ moving in an environment $E$. We say a contact $c=\{o_i, o_j\}$ has happened between two objects $o_i, o_j \in O$, when they are within a sufficiently close distance to transmit an item, i.e., when their distance is less than a threshold $d_T$. The value of $d_T$ depends on the application of interest. For example, for disease propagation through human populations $d_T$ is in the order of meters while with Bluetooth data transfer through a set of mobile devices $d_T$ is in the order of hundred meters. We call $o_i$ and $o_j$ the *contacting* objects during $c$, and we define the time interval $T_c$ within which contact persists the validity interval of $c$.

Consider a time interval $T$ during which objects in $O$ are moving in an environment $E$, and making various contacts over time. The movement of each object $o \in O$ can be modeled by the trajectory of $o$ which captures the position of $o$ at each time instant $t \in T$. We term the collection of contacts between pairs of objects in $O$ during the time interval $T$ as contact network of $O$ during $T$ and represent it by $C$. For example with Figure 1, $c_1=\{o_1, o_2\}$, $c_2=\{o_2, o_4\}$, $c_3=\{o_3, o_4\}$ and $c_4=\{o_1, o_2\}$ are the contacts occurring during $T=[0, 3]$ having validity intervals $T_{c_1}=[0, 0]$, $T_{c_2}=[1, 1]$, $T_{c_3}=[1, 2]$ and $T_{c_4}=[2, 3]$. Notice that we differentiate $c_1$ and $c_4$ although they have the same contacting objects, because by definition a validity interval is required to be continuous.

### 3.2 Reachability Query

Consider a contact network $C$ which is constructed based on the history of movement of objects $O$ in an environment $E$ during a time interval $T$. Given a pair of objects $(o_i, o_j)$, $o_i, o_j \in O$, and a time interval $T_p \subseteq T$, the reachability query $q$ verifies whether there exists a *contact path* $p_{ij}$ from $o_i$ to $o_j$ during time interval $T_p$. Intuitively, a contact path between two objects $o_i$ and $o_j$ consists of a sequence of contacts in the contact network $C$ through which any virtual item $i$ can travel the network to go from $o_i$ to $o_j$. We define a contact path from object $o_i$ to object $o_j$ as a series of contacts $(c_1, c_2, \ldots, c_n)$ in $C$, where $T_{c_i}$ overlaps $T_p$ ($1 \leq i \leq n$), and for each pair of contacts $c_i$ and $c_{i+1}$ ($1 \leq i \leq n-1$) we have 1) the contacts share an object, i.e., if $c_1=\{o_1, o_2\}$ and $c_2=\{o_3, o_4\}$ then $o_2=o_3$, and 2) $T_{c_i}$ starts before $T_{c_{i+1}}$ in time.

We call $o_i$, $o_j$ and $T_p$, query source, query destination and query interval, respectively, and denote such a query by $q : o_i \stackrel{T_p}{\leadsto} o_j$.

## 4. REACHGRID

To evaluate $q: o_i \stackrel{T_p}{\leadsto} o_j$, one approach is to first materialize the contact network $C'$, which captures all contacts that have occurred during $T_p$. It is obvious that other contacts are irrelevant to processing $q$. One can construct $C'$ as follows. Suppose trajectory of an object $o_i \in O$ during $T$ is represented by $r_i=\{(\vec{v_1}, t_1), \ldots, (\vec{v_n}, t_n)\}$ which is a sequence of position-vector and time stamp pairs $(\vec{v_j}, t_j)$, where $\vec{v_j}$ is the position vector of $o_i$ at time $t_j \in T$. Accordingly, a segment $r_i(w)$ of a trajectory $r_i$ during a time window $w$ is defined as a subset of $(\vec{v_j}, t_j)$ pairs from $r$ whose timestamps belong to $w$, i.e., $r_i(w)=\{(\vec{v_j}, t_j)|t_j \in w\}$. Assume that the set of trajectories segments from all moving objects $o \in O$ during $T_p$ is denoted by $R(T_p)$, i.e., $R(T_p)=\{r_i(T_p)\}$. A window trajectory join between two sets of trajectories $P$ and $Q$, denoted by $P \bowtie_{d_T} Q$, returns tuples $(p, q, w)$ where $p \in P$ and $q \in Q$ are within the distance of $d_T$ during $w$. $C'$ can be constructed by performing a self spatiotemporal join on $R(T_p)$, i.e., $R(T_p) \bowtie_{d_T} R(T_p)$, and subsequently creating a contact between object $o_i$ and $o_j$ at time $t$ if the join result includes $(o_i, o_j, w)$ where $t \in w$. Once generated, $C'$ can be traversed to identify any existing contact path between $o_i$ and $o_j$.

Although the aforementioned approach correctly answers reachability queries, it can be very inefficient due to redundant processing. In particular, one may not need to consider all the contacts in $C'$ to process a query $q$ in two cases. First, all contacts between objects which are not reachable from query source $o_i$ during query interval $T_p$ are irrelevant to $q$. For example for Figure 1 and $q:o_1 \stackrel{[2,3]}{\leadsto} o_2$, it is unnecessary to process the contact between $o_3$ and $o_4$ as neither can possibly be reachable from $o_1$ during $[2, 3]$. Second, we observe that $o_j$ may be reachable from $o_i$ during $T_p' \subset T_p$ where $|T_p'| \ll |T_p|$. In this case, the contacts whose validity time interval do not overlap $T_p'$ are irrelevant to $q$ and redundant for query processing. For example for Figure 1 and $q : o_1 \stackrel{[0,3]}{\leadsto} o_4$, there is no need to process the contacts occurring during [2, 3] as $o_4$ is reachable from $o_1$ during [0, 1].

Inspired by the aforementioned observations, we introduce an efficient query processing approach that given a reachability query $q$ tries to only construct the portion of $C'$ which is necessary for processing $q$. To this end, first during an offline phase we construct a spatiotemporal index structure, dubbed ReachGrid. ReachGrid enables pruning most of the contacts irrelevant to the query $q$. During the online processing phase, we incrementally find the objects reachable from the query source in the order of becoming reachable from query source when sweeping over query interval. We stop the process either if query destination is discovered reachable from the query source, or all the contacts occurring during query interval and between objects reachable from query source are processed.

## 4.1 Index Construction

ReachGrid leverages the locality of objects over space and time to avoid traversing irrelevant contacts to a reachability query. It leverages temporal locality to stop query processing as soon as a contact path between query source and destination is discovered when traversing the contacts ordered by their occurrence time. To this end, the object trajectories segments are grouped based on the time stamp of the position-vector pairs in the objects trajectories. A contact between two objects occurs when they are in close proximity. Therefore, grouping the objects based on spatial locality tends to aggregate the objects, which are in contact over time, together and in a same group. This enables traversing a subset of groups which includes only the objects reachable from query source when processing the query. ReachGrid enables temporal and spatial locality by imposing two grids on the objects trajectories. The first grid partitions the time interval $T$ ($T$ is the the time interval during which all the contacts in $C$ occurred). The second grid spatially partitions the trajectories segments within each time interval in $T$.

We construct ReachGrid as follows. First, we partition the time interval $T$ into a set of disjoint time intervals, i.e., $T=(T_1,\ldots,T_n)$. Next, we spatially partition the trajectories segments during each $T_i$, the trajectories segments in $R(T_i)$, based on locality. To this end, for each time interval $T_i$ we impose a grid $C_i$ on the environment $E$ which subsequently partitions the trajectory segments in $R(T_i)$. In this way, a grid cell $c$ in $C_i$ includes trajectories segment which span the area represented by $c$. Notice that a trajectory segment $r_i(t_i) \in R(T_i)$ may span multiple cells of $C_i$. The temporal and spatial grids' resolutions depend on the input contact network and query workload and we select them empirically in Section 6.

An example for constructed index is shown in Figure 2 where $T$ is partitioned into six time intervals. Furthermore, a $4 \times 4$ grid imposed on the environment to spatially partition the trajectories segments during $T_0$ and $T_1$. $T_0$ and $T_1$ have three and two time instances, respectively. The grid cells for the first two time intervals, i.e., the grids in $C_0$ and $C_1$, are shown while the rest are not shown for illustration purposes. Three different objects are in $O$ and represented by circle, square and triangle over time.

As the query processing progress by exploring trajectory segments in spatial grid cells, we propose to place the trajectories in a cell $c$ in $C_i$ on consecutive blocks on disk to enable efficient retrieval of necessary trajectories segments during query processing. Moreover, the position-vector and time stamp pairs $(\overrightarrow{v}, t)$ of trajectories segments in $c$ are placed on disk ordered by their time stamps. This enables avoiding processing all the trajectories segments within $c$ as soon as a contact path between query source and destination is discovered. Accordingly, placement of the cells in different time grids on disk, i.e., cells in $C_i$ versus the cells in $C_j$ where $i < j$, should be decided. Based on the same goal of early query processing termination, we place the cells in $C_i$ before the cells in $C_j$ on disk.

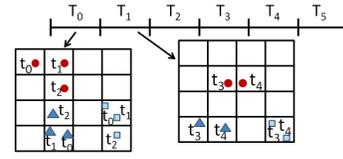
Figure 2: ReachGrid Index Example

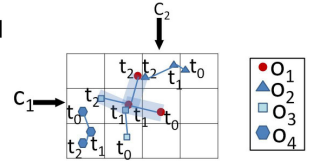
Figure 3: ReachGrid Query Processing

## 4.2 Query Processing

Query processing aims to incrementally find the objects reachable from query source by sweeping the query interval. To this end, at the beginning of the query processing, the query interval is broken into a subset of time intervals by imposing the temporal grid constructed in the previous section, i.e., $T_p=(T_j,\ldots,T_k)$. Afterward, the grid cell $c$ in $C_j$ which includes the query source at the beginning of query interval is located, i.e., cell $c$ includes the query source position at the beginning of query interval. This can be be executed in constant number of IOs assuming that an external hash table maps each object to its trajectory over time. Assume we call the set of objects reachable from query source during query processing, the *seed* set. Initially, the seed set includes only the query source. To process the reachability query, the algorithm iterates over each $T_i$ in $T_p$ and discovers new seeds. To this end, at the beginning of $T_i$ the grid cells which include the current seeds are located. Subsequently, objects which are reachable from at least one of the seeds during $T_i$ are found and added to the seed set. Notice that as soon as a new object reachable from query source is discovered, it is added to the seed set and hence the process continues with the updated seed set. The order in which new seeds are discovered is based on the time order they become reachable from any of the current seeds. In some cases, $T_j$ may be an interval whose start point different from the query interval start point. In these cases, we start processing $T_j$ from the query interval starting point. We stop the query processing if the query destination is added to seed set or when the entire query interval is processed.

The main step in the query processing is discovering new seeds during each $T_i, j \leq i \leq k$. Assume that the set of current seeds at the beginning of $T_i$ is $S_i$. The goal is to discover $S_{i+1}$, i.e., the set of seeds at the beginning of $T_{i+1}$ which is the same as that of end of $T_i$. Presume the set of grid cells in which the seeds in $S_i$ are located is denoted by $C_{S_i}$. We first discover all the other cells which may contain an object $o$ in contact with a seed during $T_i$. We call such cells potential seeds cells and denote them by $N_i$. The cells within $N_i$ can be found efficiently by creating the minimum bounding regions (MBR) of the trajectories segments of objects in $S_i$ and consequently finding and filtering the cells which are at the distance of maximum $d_T$ from those MBRs. During the query processing, whenever $N_i$ is updated, the first object $o'$ in $N_i$ is discovered which is not in $S_i$ but becomes reachable from any of seeds in $S_i$. Intuitively, we propagate a virtual item $i$ from the objects in the seed set at the beginning of $T_i$ and find the first object which receives $i$. This can be done by performing spatiotemporal join which works by sweeping time during the join interval. Consequently, we add $o'$ to $S_i$ and accordingly find $N_i$. Assume that $o'$ is discovered reachable form a seed during $[t_1, t']$ ($T_i=[t_1, t_2]$). We continue the process recursively with the updated sets but during $[t', t_2]$. Notice that during $T_i$, the retrieved cells are buffered to

851

prevent unnecessary future retrievals from disk and are discarded at the end of $T_i$.

An example is shown in Figure 3 for query processing during a $T_i$. The objects $o_1, o_2, o_3$ and $o_4$ locations at time instance $t_0, t_1$ and $t_2$, $t_0 < t_1 < t_2$, during $T_i$ are highlighted. The trajectories for each object are shown by links connecting positions at the aforementioned time stamps. Assume the query source and destination are $o_1$ and $o_2$, respectively, and query interval is $[t_0, t_2]$. The shaded area around the trajectory segment of $o_i$ denotes the MBR of the trajectory segment with the width of $d_T$. This MBR shows that $o$ is in the seed set $S_i$ and any other object whose trajectory is within the MBR of the trajectory segment of $o$ will make a contact with $o$ and be added to $S_i$. At $t_0$, $S_i$ contains $o_1$. At $t_1$, $o_1$ and $o_3$ make a contact and hence $o_3$ is added to $S_i$. During $[t_1, t_2]$ both $o_1$ and $o_3$ are in $S_i$. Finally, at $t_2$ the cells $c_1$ and $c_2$ in which $o_2$ and $o_4$ are located, respectively, are added to $N_i$ and subsequently, $o_2$ is added to $S_i$. Therefore, during $[t_0, t_2]$ query destination is reachable from query source. Due to illustration purposes, we only discussed how $N_i$ changes at $t_2$ in this example.

The entire online processing step is summarized in Algorithm 1. The algorithm gets query source, destination, interval and the index constructed during the offline process. First, query interval is quantized into time intervals from $T$. Afterward, the initialization is performed in lines 2-5. The algorithm iterates over $T_i$ in $T_p$ and for each $T_i$ it performs a join in line 9 to find the first object reachable from a seed during the interval $w$. $R_{C_{S_j}}(w)$ denotes the set of object trajectories segments during $w$ which span the cells in $C_{S_j}$. We adopt the join approach in [1] which sweeps the time interval $w$ and terminates whenever a new object, not in the seed set and reachable from query source, is discovered. Consequently, the sets are updated in line 10. Finally, the algorithm terminates when $o_j$ is added to the seeds set or all the intervals in $T_p$ are processed.

Assume each cell of $C_j$ includes the trajectories of $n_c$ distinct objects on average and each disk block contains $b_c$ cells of $C_j$ on average. Finally, assume $T'_p=[t_1, t] \subseteq T_p=[t_1, t_2]$ is the smallest time interval during which query destination is reachable from source. If query destination is not reachable from query source during $T_p$, we assume $T'_p=T_p$. The following theorem proves the complexity of ReachGrid query processing and index construction.

THEOREM 4.1. *ReachGrid can be constructed with $O(|O||T|)$ IOs. The IO complexity of query processing is $O(\frac{|O||T'_p|}{n_c \times b_c})$.*

We skip the details of the proof due to lack of space.

**Algorithm 1** Query Processing
1: **procedure** QUERY PROCESSING($o_i$, $o_j$, $T_p$, $I$)
2:    $T_p=(T_j, T_{j+1}, \ldots, T_k)$
3:    $S_j=o_i$                                            ▷ Initializing the seed set
4:    $C_{S_j}$=FindCells($S_j, t$)        ▷ Find the cells containing the seed
5:    $C_{S_j}$=Update($C_{S_j}, N_j$)    ▷ Update $C_{S_j}$ based on the cells in $N_j$
6:    **for** $i=j$ to $k$ **do**
7:       $w=T_i=[t_1, t_2]$
8:       **repeat**
9:          $(o', t') = R_{C_{S_i}}(w) \bowtie_{d_T} R_{C_{S_i}}(w) \triangleright o' \notin S_i$ and is reachable from a seed
10:          $w=[t', t_2]$
11:          Update $N_i$, $C_{S_i}$ and $S_i$
12:       **until** $o' = $ NULL or $o' = o_j$        ▷ Termination condition
13:       **if** $o_j \in S_i$ **then**
14:          Return 'reachable'
15:       **end if**
16:    **end for**
17:    Return 'not reachable'
18: **end procedure**

## 5. REACHGRAPH

In this section, we first present ReachGraph index construction steps and thereafter discuss ReachGraph query processing.

### 5.1 Index Construction

To construct the ReachGraph for a given contact network $C$, we start from $C$ and apply a series of transformations to $C$ that eventually converts it to the ReachGraph hyper graph $H_N$. The transformations are performed in two phases, namely *reduction* phase and *augmentation* phase. First, we observe that in a contact network $C$ one can identify disjoint subset of nodes, where all nodes in a subset are equivalently reachable or not reachable to/from any other node $v$ in $C$. Accordingly, at the reduction phase we precompute these subsets and reduce all nodes in each subset (along with their connections) to a single hyper node. We call the resulting hyper graph $D_N$ which is a significantly reduced version of $C$ in size. Next, at the augmentation phase, to further improve ReachGraph we precompute the reachability between pairs of nodes in $D_N$ at predefined time intervals. We perform this precomputation at several time resolutions and accordingly augment $D_N$ with a hierarchy of extra links to generate the ReachGraph hyper graph $H_N$. With $H_N$, a reachability query can be effectively broken into a set of precomputed reachability queries for real-time query answering.

There are two principles in disk placement of $H_N$ vertices which can improve the query processing. First, an efficient placement should place vertices which are reachable to each other on a same disk block. In this way, while retrieving a vertex during the query processing, a set of vertices which should retrieved in the future are read and buffered as well. Second, there is an order inherited in how the vertices of $H_N$ are traversed during query processing which should be leveraged when storing $H_N$ on disk. This ordering is enforced by the time order at which the contacts in the vertices of $H_N$ are occurred. We explain how to consider these two principles in storing $H_N$ on disk to enable efficient query processing.

In the rest of this section, we first present our model for $C$ as a so-called time expanded network. Next, we explain the aforementioned transformations $C \xrightarrow{\text{Reduction}} D_N$ and $D_N \xrightarrow{\text{Augmentation}} H_N$ in detail. Finally, we discuss how to store $H_N$ on disk.

#### 5.1.1 Contact Network Model

We represent a contact network $C$ with Time Expanded Network (TEN) model [14]. TEN captures the time dependency of a network by including a separate instance of the network at each time instance. Accordingly, each object $o_i$ at time instance $t \in T$ is associated with a separate vertex $o_i(t)$. To capture contacts, a bidirectional edge $e=(o_i(t), o_j(t))$ is introduced between $o_i(t)$ and $o_j(t)$ if they are in contact at time $t$. Such an edge captures the fact that an item can transfer from $o_i$ to $o_j$ at $t$. Note that we assume transfer delay is negligible and hence, $e$ is bidirectional. Moreover, an edge is introduced between vertices corresponding to the same object at consecutive time instances, i.e., an edge $e'=(o_i(t), o_i(t+1))$ is created between $o_i(t)$ and $o_i(t+1)$ at each time $t$. In this case, $e'$ is a directional edge which shows that $o_i$ can hold an item during $[t, t+1]$. We define a graph $G_t$ of all vertices and edges at time $t$, i.e., $G_t=(V, E)$ where $V=\{o_i(t)|o_i \subseteq O\}$, as a snapshot of $C$ at $t$.

Figure 4 (a) shows an example $C$ which corresponds to the contact network in Figure 1. With $G_0$ in Figure 4 (a), $V=\{o_1(0), o_2(0), o_3(0), o_4(0)\}$ and $E=(o_1(0), o_2(0))$. It is easy to observe that $o_j$ is reachable from $o_i$ during $T_p=[t_1, t_2]$ if and only if there is a path from $o_i(t_1)$ to $o_j(t_2)$. This path is representing the contact path from $o_i$ to $o_j$ during $T_p$. For example, in Figure 4 (a), $o_4$ is reachable from $o_1$ during $T_p=[0, 1]$ given the path $(o_1(0), o_2(0), o_2(1), o_4(1))$.

852

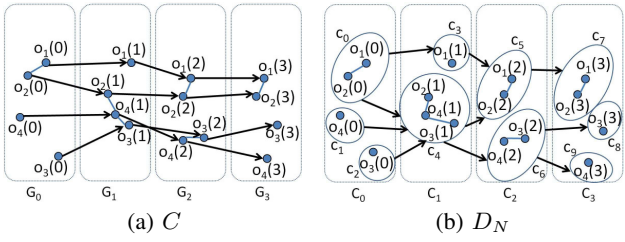

(a) $C$

(b) $D_N$

**Figure 4: TEN model of $C$ (a) and the corresponding DAG (b)**

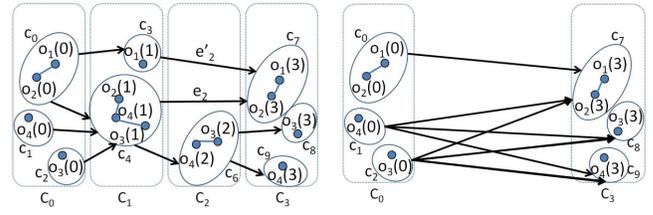

**Figure 5:** $D_N$ at the end of reduction step

**Figure 6:** $D_{N_3}$ for $H_N$ whose $D_{N_1}$ is the graph in Figure 5

### 5.1.2 Transforming the Contact Network

#### 5.1.2.1 Transforming by Reduction.

In the reduction phase, we perform two distinct steps to convert $C$ into a hypergraph $D_N$ with significantly smaller size. Reducing $C$ makes it more efficient to traverse for finding possible contact paths during query processing. Notice that these reduction steps are lossless and preserve the accuracy of query processing. we first state two properties which are utilized for reduction.

PROPERTY 5.1. *[Snapshot Symmetry] If $o_j$ is reachable from $o_i$ during a time instance $t$, i.e., query interval $T_p=[t,t]$, $o_i$ is reachable from $o_j$ at the same interval.*

PROPERTY 5.2. *[Transitivity] Suppose $o_j$ is reachable from $o_i$ during $T_p=[t_1,t_2]$ and $o_k$ is reachable from $o_j$ during $T'_p=[t'_1,t'_2]$. If $t_2 \leq t'_2$ then $o_k$ is reachable from $o_i$ during $T''_p=[t_1, t'_2]$.*

At the first step of the reduction phase, the idea is to precompute and materialize the reachability between objects at each time instance $t$. According to properties 5.1 and 5.2, the connected components of $C$ capture the set of objects that are reachable from each other at $t$. For instance, in Figure 4(b), $c_4=\{o_2(1), o_3(1), o_4(1)\}$ which captures the fact that all objects $o_2, o_3$ and $o_4$ are reachable from each other at time instance $t$=1. Furthermore, if one object from a connected component $c \in G_t$ is reachable from another object in a connected competent $c' \in G_{t'}$ during $T_p=[t,t']$, then it is easy to deduct from properties 5.1 and 5.2 that all object in $c$ are reachable from all other objects in $c'$ during $T_p$. Accordingly, at the first step of the reduction phase, we transform $C$ to a graph $D_N$ whose vertices are the connected components of $C$. To this end, first in every $G_t \in C$ we replace all the vertices within the same connected component $c$ by a single vertex represented by $c$. Suppose the collection of the connected components of $G_t$ are denoted by $C_t$. Next, we create an edge from every $c \in C_t$ to every other $c' \in C_{t+1}$, if in $C$ we find at least one edge from a vertex in $c$ to a vertex in $c'$. This transforms $C$ into a directed acyclic graph (DAG) $D_N$ with significantly smaller number of vertices and edges as compared to $C$ while preserving reachability between objects. With $D_N$, $o_j$ is reachable from $o_i$ during $T_p=[t_1,t_2]$ if the connected component of $o_j(t_2)$ is reachable from the connected component of $o_i(t_1)$. Therefore to answer a reachability query, we need to find the corresponding connected components of $o_i(t_1)$ and $o_j(t_2)$ given $o_i(t_1)$ and $o_j(t_2)$ at the query time. As we explain later, we generate and use external hash table $\mathcal{H}_t$ for each time instance $t \in T$ to locate the the connected component corresponding to each vertex $o_i(t)$.

The second step of reduction phase merges identical connected components in consecutive $G_t$s over time. If a set of objects $O' \subseteq O$ are reachable from each other (and only from each other) during a time interval $T' \subseteq T$, in $D_N$ they all belong to snapshots of the same connected component during $T'$. Therefore, to further reduce the size of $D_N$ we can keep one copy of such connected component during $T'$ and consider it as the connected component of objects in $O'$ during the entire $T'$. For example, in Figure 4(b) $c_5$ and $c_7$ are snapshots of the same connected component during $T'=[3, 4]$ and can be merged. To generalize, assume a set of connected components $c_t \in C_t, c_{t+1} \in C_{t+1}, \ldots, c_{t+n} \in C_{t+n}$ all have the same members $O'$, and $T'=[t, t+n]$. In such a case, we remove $c_t, \ldots, c_{t+n-1}$ and connect parent of $c_t$ in $D_N$ (say a connected component in $G_{t-1}$ denoted by $d$) to $c_{t+n}$ by a weighted edge $e(n)$. We call $e(n)$ an *aggregated* edge where the weight captures the fact that for the next $n$ time instances, $d$ is only reachable to objects in $O'$. Figure 5 shows $D_N$ from Figure 4(b) after this step of reduction. $c_5$ is removed, $c_4$ and $c_3$ are connected to $c_7$ by aggregated edges $e(2)$ and $e'(2)$. This reduction can significantly shrink $D_N$, especially when the sampling rate for objects positions is high relevant to the objects moving speed.

#### 5.1.2.2 Transforming by Augmentation.

In order to find a path between two connected components $c_i \in C_t$ and $c_j \in C_{t'}$, we can simply expand $D_N$ starting from $c_i$ and check if we can find a path that reaches $c_j$. Although $D_N$ is much smaller than $C$, such expansion can still take a long time to terminate. Hence, we propose to precompute reachability between certain vertices of $D_N$ to enable quick traversal of $D_N$.

In particular, we propose to precompute reachability during different predefined time intervals. To this end, we break $T$ into a set of disjoint intervals $I_1, I_2, \ldots, I_n$ with equal length $L$, and precompute reachability between vertices in $C_{t_a}$ and $C_{t_b}$ for each $I_i=[t_a, t_b]$. Accordingly, $D_N$ is augmented with a new directed edge from every connected component $c \in C_{t_a}$ to every other connected component $c' \in C_{t_b}$ if there is a path of length $L$ from $c$ to $c'$. We call such edges the *long edges* and weight them by $L$ which indicates the number of time instances that encompass. The resulted augmented hyper graph $H_N$ can be considered as the union of $D_N$ with a new graph consisting of long edges each with a weight $L$. We term the latter graph contact network at the $L$-th resolution and denote it by $D_{N_L}$. Accordingly, $D_N$ can be considered as the contact network at first resolution or $D_{N_1}$. One can extend this idea and precompute reachability at other time intervals to generate a multi-resolution graph $H_N = D_N \cup D_{N_{L_1}} \cup D_{N_{L_2}} \cup \ldots \cup D_{N_{L_n}}$. However, this can significantly increase the number of edges if overdone and hence adversely reduce the efficiency of query expansion. In Section 6, we experimentally select the optimal resolutions for $H_N$. Figure 6 depicts $D_{N_3}$ where $D_N$ shown in Figure 5.

### 5.1.3 Disk Placement

We distinguish two cases in traversal of $H_N$ which is a disk-resident hyper graph. With the first case, internal memory can hold $c \times |V(H_N)|$ values where $V(H_N)$ is the set of vertices in $H_N$ and $c$ is a small constant ($c \approx 12.375$) [15]. In this case, it is possible to construct the DFS tree of the graph and maintain it in the internal memory and traverse it during query processing to verify reachability. With the second case, the aforementioned assumption on the number of vertices does not satisfy. For this case, we adopt the idea of external BFS presented in [12] to enable efficient retrieval of vertices during query processing. Similar to [12], we partition $H_N$ and place the vertices within the same partition on consecutive disk blocks. However, we adopt the technique for directed graphs as $H_N$ is a DAG. To this end, we first sort the vertices of $H_N$ in topological order which is the same order in which $H_N$ is traversed during reachability query processing. Notice that finding such order is trivial as $H_N$ vertices are created over $T$ in topological order (vertices in $C_i$ are generated before that of $C_{i+1}$). Afterward, from each vertex $v$ we find all the vertices $U$ with the shortest distance of of at most $d_p$ from $v$, i.e., vertices at the *depth* of at most $d_p$ from $v$. The set of vertices in $U \cup v$ are reachable from $v$ and forms a partition $p_v$. We term $v$ the root of $p_v$. We iterate over the vertices and create a partition rooted at a vertex $u$ if $u$ is not already assigned to a partition. Notice that only the edges in $D_N$ are considered in creating the partitions and hence long edges are ignored during the partitioning process to preserve the temporal locality of graph vertices within the same partition. The partitions are placed on disk in the same order they are generated.

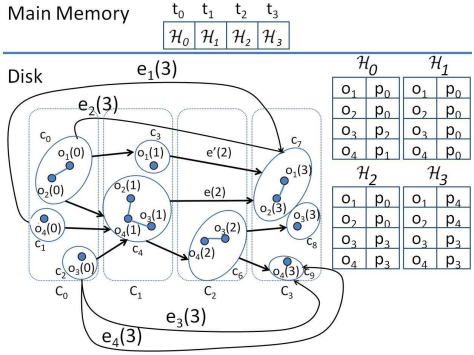

**Figure 7: ReachGraph for the contact network in Figure 5**

The final index for our running example is shown in Figure 7 where the long edges are denoted by $e_i(3), i = 1, \ldots, 4$, and the aggregated edges by $e(2)$ and $e'(2)$. The hyper graph $H_N$ and the hash tables which associate objects with the partitions of $H_N$ are located on the disk. Each hash table $\mathcal{H}_t$ locates the partition which contains $o_j(t)$ given object $o_j$ and the time instance $t$. In this example, five partitions $p_0, p_1, \ldots, p_4$ are generated where their connected component members are $\{c_0, c_3, c_4\}$, $\{c_1\}$, $\{c_2\}$, $\{c_6, c_8, c_9\}$ and $\{c_7\}$, respectively. The members of the connected components are placed within the vertices of $H_N$ as we discuss in next section. Although not shown in the figure, we store the reverse graph of $D_{N_1}$ on disk as well, i.e., if $e=(u,v) \in D_{N_1}$ then we add $e = (v,u)$ to $H_N$. This enables efficient bidirectional traversal of $H_N$ as we discuss in the next section. Finally, a hash table is stored in main memory to enable fast lookup of $\mathcal{H}_t$ for a given $t$ and consequently finding the partition of $H_N$ which includes query source (destination) at the beginning (end) of query interval on disk.

## 5.2 Query Processing

Consider a reachability query $q: o_i \stackrel{T_p}{\leadsto} o_j$ where $T_p=[t_1,t_2]$. To process $q$, one can first find the vertices $v_1$ and $v_2$ in $H_N$ which representing the connected component of $o_i$ and $o_j$ at $t_1$ and $t_2$, respectively. Afterward, starting from $v_1$, $H_N$ can be traversed either by BFS or DFS techniques to visit all the vertices at the depth of at most $|t_2-t_1|$ from $v_1$. $o_j$ is reachable from $o_i$ during $T_p$ if and only if $v_2$ is among the visited vertices. Unfortunately, this approach may visit a huge number of vertices specially when $t_1 \ll t_2$. In this section, we propose two powerful ideas which significantly reduce the number of visited vertices during $H_N$ traversal. First, we leverage multi-resolution index to traverse $H_N$. Consequently, whenever possible the long edges with the largest weights are taken during traversal (the traversal is performed on the higher resolutions first) to enable fast traversal of $H_N$. Second, motivated by transitivity property 5.2, we traverse $H_N$ from both directions to find a possible contact path between query source and destination faster. In particular, $H_N$ is traversed forward starting from query source and in parallel it is traversed backward on the reverse of $D_N$ starting from query destination. The traversal is terminated in two cases. Either, an object which is reachable from query source and reachable to query destination is found, or $H_N$ is traversed in both directions until the bidirectional traversal stops at the middle of the query interval.

Counterpart to traversal algorithm for memory-resident graphs, external graphs traversal algorithms are studied in the literature as well [12, 17]. We denote external BFS and DFS by E-DFS and E-BFS, respectively. Although both E-DFS and E-BFS can be adopted to traverse $H_N$, we adopt E-BFS to enable bidirectional traversal of $H_N$. Accordingly, our ReachGraph query processing works by performing E-BFS in parallel from $v_1$ and $v_2$ where the search from $v_2$ traverses the reverse graph of $D_{N_1}$. Assume the set of objects in vertices visited during forward traversal, i.e., traversal originating from $v_1$, is denoted by $O_F$. Accordingly, we denote the set of objects in vertices visited during backward traversal by $O_B$. The traversal is terminated either when $O_B \bigcup O_F$ becomes non-empty or when all the vertices reachable from $o_i$ during $[t_1, \frac{(t_1+t_2)}{2}]$ and reachable to $o_j$ during $[\frac{(t_1+t_2)}{2}, t_2]$ are traversed. In the first case query destination is reachable from source while this is not true for the latter case. A partition is retrieved and buffered during traversal to enable in-memory lookup of some of the future vertices. Older partitions in memory can be discarded when there is not enough space for new partitions. During forward traversal, if a vertex is connected to long edges, the edges with the largest weight are traversed and the other edges are ignored. We term this approach Bidirectional Multi-resolution BFS or BM-BFS.

The pseudocode of BM-BFS technique is presented in Algorithm 2. The algorithm first finds the vertices $v_1, v_2 \in H_N$ in lines 2-3. The function FindVertex($p, o, t$) gets a partition $p$, an object $o$ and a time instance $t$ and returns the vertex of $H_N$ which contains $o(t)$. Afterward, two queues are initialized for the forward and backward traversal of the input graph in line 4. $O_F$ and $O_B$ are also initialized in line 5. We denote the set of object whose instances are included in $v$ by $O_v$. The algorithm runs forward (line 7) and backward traversal (lines 8) in parallel by running ProcessQueue procedure until both $Q_F$ and $Q_B$ become empty or reachability is verified. With ProcessQueue procedure, the vertex $v_h$ in the head of either queue is extracted in line 2. Each object in $O_{v_h}$ is examined to check whether it is already visited in the reverse traversal (lines 5-8). If this is the case, query destination is reported reachable from query source. Afterward, each children $v$ of $v_h$ is added to the traversal queue to enable the next steps of traversal. Child(v,direction) procedure returns the edges at the highest resolution originating from $v$ whose end points are the vertices representing time instance $t \in [t_1, \frac{t_1+t_2}{2}]$ and $t \in [\frac{t_1+t_2}{2}, t_2]$ for forward and backward traversal directions, respectively. The following proves the correctness of BM-BFS.

**Algorithm 2** BM-BFS

```
1:  procedure BM-BFS(o_i, o_j, T_p = [t_1, t_2], H_N)
2:     v_1 = FindVertex(H_{t_1}(o_i), o_i, t_1)
3:     v_2 = FindVertex(H_{t_2}(o_j), o_j, t_2)
4:     Q_F.push(v_1), Q_B.push(v_2)
5:     O_F.add(O_{v_1}), O_B.add(O_{v_2})
6:     while !Q_F.isEmpty()||!Q_B.isEmpty() do
7:        ProcessQueue(O_B,Q_F,F)
8:        ProcessQueue(O_F,Q_B,B)
9:     end while
10:    return false
11: end procedure
1:  procedure PROCESSQUEUE(O,Q,direction)
2:     if !Q.isEmpty and v_h = Q.pop() is not visited before then
3:        for o ∈ O_{v_h} do
4:           if O.contains(o) then
5:              return true
6:           end if
7:        end for
8:        for c ∈ Child(v_h,direction) do
9:           Q.add(c)
10:       end for
11:    end if
12: end procedure
```

THEOREM 5.3. *BM-BFS verifies the reachability from query source to destination during query interval.*

PROOF. First, assume that $H_N$ only includes one resolution, i.e., $H_N = D_{N_1}$. $H_N$ is a DAG whose vertices are topologically sorted and time stamped. The forward traversal visits all the vertices representing contacts with validity interval subset of $[t_1, \frac{t_1+t_2}{2}]$ and reachable from query source. Accordingly, the backward traversal visits all the vertices representing contacts with validity interval subset of $[\frac{t_1+t_2}{2}, t_2]$. Therefore, if a path $p$ from $v_1$ to $v_2$ exists, then the vertices in $p$ are discovered after forward and backward traversal of $H_N$. In addition, the vertices in $p$ are time stamped and therefore, the order of vertices in $p$ are preserved during traversal of $H_N$. When we consider long edges during traversal, some vertices of $H_N$ which representing specific time instances may not be visited. However, general connectivity of the graph is preserved at all the resolutions and therefore by taking long edges the query can be still verified correctly. Also, based on the transitive property 5.2 the early termination condition accurately terminates the traversal. This completes the proof. □

Assume that each partition includes instances of $n_p$ distinct objects and each disk block holds $b_p$ partitions on average. The following theorem proves the complexity of ReachGraph query processing and index construction ($|T'_p|$ is defined in Theorem 4.1).

THEOREM 5.4. *The ReachGraph index can be constructed with $O(|O||T|)$ IOs. The query processing IO complexity is $O(\frac{|O||T'_p|}{n_p \times b_p})$.*

We skip the details of the proof due to lack of space.

GRAIL [18] is one of the most efficient graph reachability approaches for memory resident graphs. It works based on the idea of randomized interval labeling of graph vertices. Table 1 compares the index construction and query time complexity of Reach-Grid and ReachGraph with that of GRAIL when adopted on disk-resident $D_N$ to process reachability queries. Our approaches significantly outperforms GRAIL because of efficient disk placement and also early termination of queries ($|T'_p| \leq |T_p|$). With GRAIL, $d$ is a small constant and it is the number of intervals assigned to each graph vertex. $n_r$ is the average number of objects which are reachable from any object $o \in O$ at each time instance $t \in T$.

|  | GRAIL | ReachGraph | ReachGrid |
|---|---|---|---|
| Query Time | $O(|O||T_p|n_r)$ | $O(\frac{|O||T'_p|}{n_p \times b_p})$ | $O(\frac{|O||T'_p|}{n_c \times b_c})$ |
| Construction Time | $O(d|O||T|)$ | $O(|O||T|)$ | $O(|O||T|)$ |

**Table 1: Complexity Comparison**

| Dataset | Size |
|---|---|
| $RWP_{10k}$ | $190GB$ |
| $RWP_{20k}$ | $380GB$ |
| $RWP_{40k}$ | $760GB$ |
| $VN_{1k}$ | $23GB$ |
| $VN_{2k}$ | $46GB$ |
| $VN_{4k}$ | $92GB$ |

| Memory Size | $4GB$ |
|---|---|
| Disk Size | 5 disks each 1.36 TB |
| OS | Windows 7 SP1 64-bit |
| CPU | 3.34GHz |
| Page Size | 4kb |

**Table 2: Data Collection Size**   **Table 3: System Specifications**

## 6. EXPERIMENTS

We perform our experiments on both synthetic and real datasets modeling the contacts between moving objects which are either vehicles or individuals. Our synthetic data sets are generated by two different data generators. The first data generator, GMSF [3], models the movement of individuals in an environment of $100km^2$ assuming their movement patterns follow random waypoint model with the average speed of $2m/s$. The trajectories samples are captured every 6 seconds. Random waypoint is one of the most used models in literature to model individuals' movement. With this model, every individual selects a random destination and speed and then moves toward that destination. Afterward, she selects another random destination and moves toward it [11]. The second data generator is the Brinkoff generator which is commonly used for generating realistic moving objects [4]. We generated the trajectories of a constant set of vehicles moving on the road network in San Francisco city covering an area of approximately $300km^2$. The vehicles locations are recorded on average every 5 seconds. The reason of using two different synthetic data generators is to study the difference between the case of reachability query processing for different categories of moving objects, i.e., individuals and vehicles. In particular, vehicles are restricted to move on a road network while individuals can move to any environment point. With the first synthetic data generator we generate 1000, 2000 and 4000 vehicles trajectories. We denote these datasets by $VN_{1k}$, $VN_{2k}$ and $VN_{4k}$, respectively, and term the collection, VN datasets. With the second synthetic data generator, we generate 10,000, 20,000 and 40,000 individuals' trajectories. We term these datasets $RWP_{10k}$, $RWP_{20k}$ and $RWP_{40k}$, respectively, and call the set of these datasets, RWP datasets. The reason of generating more objects trajectories with the second dataset is that the objects are distributed in the entire space with the second generator as opposed to the first generator in which objects only move on the road network. With both generators, we generate trajectories for the duration of four months (more than 119 days). Accordingly, RWP and VN datasets include more than 1,700,000 and 2,048,000 time instances, respectively. The size of the data for each dataset is represented in Table 2.

Our real dataset captures the movements of vehicles in the city of Beijing. This dataset covers the GPS tracks of more than 2500 distinct vehicles collected during a day. The vehicles GPS tracks cover an area of approximately $600km^2$. The vehicles locations are recorded every minute and further interpolated to reflect the locations for every five seconds. Unfortunately, because of the small scale of this datasets we only use it in a subset of experiments.

Our experimental system specification is presented in Table 3. For each experiment setting, we run the algorithm 400 times to compute the average values. The query sources, destinations are selected randomly and query interval is selected as a random interval where the length of the interval is a random number between 150 and 350 unless otherwise stated. We presume vehicles are contacting each other by communicating over DSRC protocol which has the effective range of 300 meters. Accordingly, we assume individuals are making contacts by communicating over Bluetooth protocol which has the typical range of 25 meters. Therefore, we set $d_T$=25 for RWP and $d_T$=300 for VN datasets.

Finally, to measure the performance of reachability query processing we measure the number of random IOs. Hence, the sequential IOs are normalized to random accesses by assuming that each random access costs as much as 20 sequential accesses [6]. Notice that these numbers are system dependent, however, the general trends in the results should be obtained for machines with the different settings as well. The rest of this section is organized as follows. We first evaluate the efficiency of ReachGrid and ReachGraph approaches, respectively. Thereafter, we present the empirical comparison between ReachGrid and ReachGraph. Finally, we compare our approaches with existing graph reachability algorithms.

## 6.1 ReachGrid

In this section, we first focus on the efficiency of the index construction and then query processing step of our ReachGrid.

### 6.1.1 Index Construction

The performance of the ReachGrid depends on the resolution of temporal and spatial grids which quantize time interval $T$ and environment $E$, accordingly. There is a tradeoff in selecting both temporal and spatial resolutions. By increasing any of the resolutions, the number of random accesses to disk blocks increases when processing a reachability query and hence the number of IOs increases. The reason is that the locality in time and space is not fully leveraged. On the other hand, decreasing the resolution of grids results in placement of huge number of trajectory segments within a grid cell. As the result, many trajectory segments which are irrelevant for query processing are processed for each reachability query. This increases the number of IOs during query processing.

Here, we empirically optimize the grids resolutions by varying both temporal and spatial grids and selecting a combination which minimizes the number of IOs when processing reachability queries. There are huge possible number of values for the the combination of temporal and spatial resolutions, and therefore, we assume the same resolution for all the spatial grids $C_i$ to reduce the number of possible combinations. We vary temporal resolution from 5 to 80 for both datasets and spatial resolution from $128m$ to $10km$ ($17km$) for RWP (VN) datasets and select a combination which minimizes the number of IOs while processing reachability queries. We denote the optimal spatial and temporal resolutions by $R_S$ and $R_T$, respectively. With RWP datasets, $R_S$=1024m and $R_T$=20 and accordingly, with VN datasets, $R_S$=17km and $R_T$=20. With VN datasets, the optimal ReachGrid indexes have lower resolutions than that of RWP datasets. The reason is that VN datasets capture the movement of fewer objects as compared to RWP datasets and hence spatial grids are larger to place more objects within the same cell. Figures 8 (a) and (b) show how IO count varies when temporal and spatial resolutions vary for RWP datasets, respectively. With Figure 8 (a) temporal resolution is 20 and with Figure 8 (b) the spatial resolution equals $1024m$. Because of lack of space and the fact that VN datasets results also follow the same pattern, we do not show the results for VN datasets.

We also measured the time required to construct the optimal ReachGrid indices. The results are shown in Figures 9 (a) and (b) for RWP and VN datasets. The x-axis shows the length of time pe-riod $T$ over which ReachGrid index is constructed. All these intervals share the same starting point but different ending point. Over all the cases, the index construction time is less than 4.3 hours. As expected, increasing the number of objects and duration of $T$ makes index construction slower.

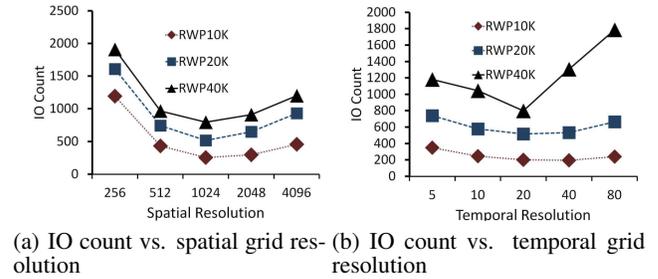

(a) IO count vs. spatial grid resolution  (b) IO count vs. temporal grid resolution

**Figure 8: ReachGrid resolutions optimization**

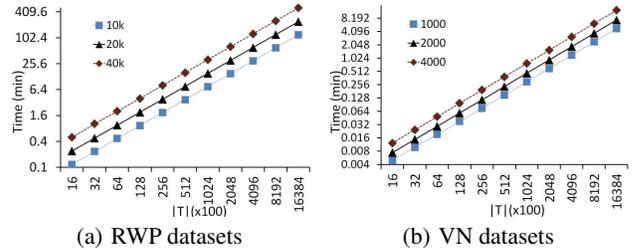

(a) RWP datasets  (b) VN datasets

**Figure 9: ReachGrid construction time**

### 6.1.2 Query Processing

To evaluate the efficiency of online ReachGrid query processing, we compare ReachGrid and naïve approach, termed SPJ, which generates the contact network $C'$ relevant to query interval on the fly and afterward traverse it to verify reachability between query source and destination. SPJ generates $C'$ by retrieving all the trajectories segments which overlap with the query interval. Based on our experiments, our ReachGrid approach outperforms SPJ by at least 96% for all RWP and VN datasets. The reason is that our ReachGrid online query processing algorithm avoids constructing the portion of contact network which is irrelevant for query processing by intelligent traversal of the contact network.

## 6.2 ReachGraph

Here, we first study the efficiency of index construction and afterward the online query processing approaches of ReachGraph.

### 6.2.1 Index Construction

In this section, we first focus on evaluating the efficiency of index construction for the basic contact network ($D_N$) and afterward, evaluate the efficiency of the augmentation step. We conclude this section by studying the placement of ReachGraph on disk.

#### 6.2.1.1 Contact Network Size.

Here, we empirically measure the contact network size by counting the number of vertices ($|V|$) and edges ($|E|$) of contact network ($D_N$) when generating contact network for different time intervals $T$. The results for RWP datasets are shown in Figures 10 (a) and (b) for edges and vertices, respectively. The results for VN datasets follow the similar pattern and omitted due to space constraints. The x-axis represents the length of time interval $T$ during which the

856

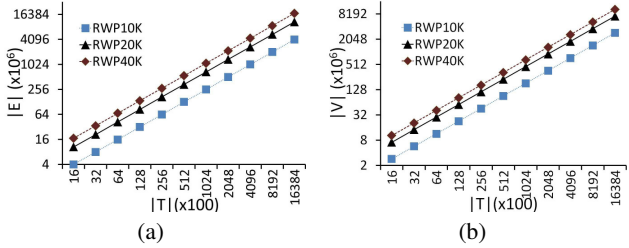

**Figure 10: Contact network edges (a) and vertices (b)**

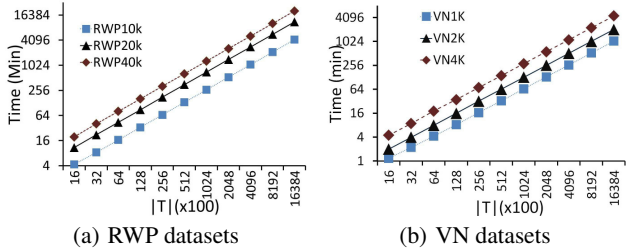

(a) RWP datasets  (b) VN datasets

**Figure 11: Contact network ($D_N$) construction time**

contact network is constructed assuming that all time intervals are starting from the same time instance, i.e., $T=[0,|T|]$. As expected, $|E|$ and $|V|$ increases when $|T|$ increases. The reason is that the number of contacts and accordingly the number of edges increase when $|T|$ increases. Accordingly, the number of edges and vertices increases when the number of objects increases as well. The most important observation from this experiment is that the contact network can become prohibitively large to reside in the main memory. In particular, the number of edges and vertices are more than 17,466 and 10,545 million for $RWP_{40k}$, respectively.

Next, we measure the efficiency of the reduction step proposed in Section 5. To this end, we compare the number of vertices and edges of $C_N$ and that of $D_N$ for the same settings of the experiments in this section. With RWP datasets, over all the cases on average the number of vertices (edges) of $D_N$ are 81% (80%) less than that of $C_N$, respectively. Similarly, with VN datasets, over all the cases the number of vertices (edges) of $D_N$ are 64% (61%) less than that of $C_N$. The results show that reduction step can significantly reduce the size of contact network represented in TEN.

#### 6.2.1.2 Contact Network Construction Time.

In this section, we measure the construction time of $D_N$ for different time intervals $T$. The results are shown in Figures 11 (a) and (b) and for RWP and VN datasets. For all datasets, increasing the number of objects and $|T|$ increases the construction time. The reason is that more contacts needs to be processed in order to create the contact network. With our experimental setting, the construction time for all datasets is less than 14 days. Although this running time is large, it reflects the time it takes to construct the entire contact network over $T$. However, it is also possible to construct the contact network incrementally over time by acquiring the objects positions at new time instances and appending corresponding new vertices and edges to the previously constructed contact network.

#### 6.2.1.3 Multi-resolution Graph.

In this section, we study the performance of constructing the contact network at various resolutions. To this end, we mea-sure the average degree of vertices of $H_N$ at different resolutions ($D_{N_2}, D_{N_4}, \ldots, D_{N_{32}}$). The average degree for $D_{N_i}$ only considers vertices which have at least one edge at $N_i$th resolution. Table 4 shows the results for $RWP_{40K}$ and $VN_{4k}$ which have the largest number of objects among RWP and VN datasets and also $VN_R$ which corresponds to our real dataset. As the contact network resolution increases, the average degree of vertices in the corresponding resolution increases. The reason is that over larger time intervals, objects are reachable from more objects and hence more long edges are introduced at higher resolutions. $VN_R$ has significantly smaller average vertex degree than the other datasets. The reason is that the size of contact network $D_N$ for this dataset is much smaller than that of other datasets. We decide the optimal number of Reach-Graph resolutions in Section 6.2.1.4.

| Resolution | $VN_{4k}$ | $RWP_{40k}$ | $VN_R$ |
|---|---|---|---|
| $D_{N_2}$ | 2.9 | 3.0 | 1.5 |
| $D_{N_4}$ | 6.1 | 8.1 | 1.7 |
| $D_{N_8}$ | 16.3 | 33.4 | 2.3 |
| $D_{N_{16}}$ | 55.5 | 75.6 | 3.69 |
| $D_{N_{32}}$ | 221.4 | 322 | 9.0 |

**Table 4: Average vertex degree for $D_{N_i}$**

#### 6.2.1.4 Disk Placement.

Here, we empirically optimize the placement of multi-resolution contact network graph on disk. ReachGraph has two parameters, i.e., the number of resolutions and the depth of partitioning, which needs to be optimized in order to construct and place the index on disk. Here, we empirically find the optimal values for both parameters. To this end, we vary partitions depths from 1 to 64 and the number of resolutions from 1 to 7 and count the number of IOs for both datasets. Based on our experiments, the optimal partitions depth and the number of resolutions are 32 and 6, respectively, i.e., $d_p$=32 and $H_N=D_{N_1} \cup D_{N_2} \cup \ldots \cup D_{N_{32}}$.

Figure 12 shows how changing the depth of partitions varies the number of IOs for $RWP_{20k}$ and $VN_{2k}$ datasets when processing reachability queries ($H_N$ includes contact network at the first six resolutions). Increasing the depth of partitions gives the opportunity to buffer more vertices which will be visited in the future and hence reduces the total number of IOs. On the other hand, if the partitions become too large then many vertices redundant for query processing are retrieved from disk which will deteriorate the performance of query processing. Therefore, there is a trade-off between partitions depth and IOs count. Similar trade-off is present between the number of ReachGraph resolutions and IOs count.

### 6.2.2 Query Processing

Here, we evaluate the efficiency of online ReachGraph query processing. The goal of this experiment is to study how bidirectional traversal and multi-resolution index construction techniques improve the performance of ReachGraph. To this end, we compare the efficiency of bidirectional multi-resolution traversal (BM-BFS) approach with bidirectional traversal (B-BFS) and external DFS (E-DFS) approaches. B-BFS traverses $H_N$ similar to BM-BFS but only at the single resolution of $D_N$. E-DFS is the naïve approach which only checks whether there is a path on $H_N$ from query source to destination during query interval. We select E-DFS as the baseline approach as it is faster than E-BFS. Notice that E-DFS does not investigate the members of the connected components as opposed to BM-BFS and B-BFS and therefore it only finds the contact paths with the length of query time interval. The results for $RWP_{20k}$ and $VN_{2k}$ are shown in Figure 13. BM-BFS is outperforming E-DFS and B-BFS for more than 80% and 15%, respectively, for both datasets. The reason is that it leverages long edges to make traversal faster and at the same time investigates the objects within connected components to stop the traversal as soon as a contact path is found between query source and destination. B-BFS also outperforms E-DFS significantly because of terminating graph traversal as soon as a contact path is discovered between query source and destination.

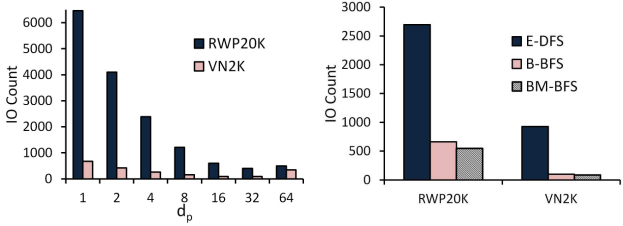

**Figure 12: IO count vs different partition depths**

**Figure 13: ReachGraph online query processing for different approaches**

## 6.3 ReachGrid vs. ReachGraph

In this section, we compare the efficiency of ReachGrid and ReachGraph. We generate random queries with varying queries intervals of 100, 300 and 500 time instances and compare the number of IOs for ReachGrid and ReachGraph (BM-BFS) approaches. The results are shown in Figure 14 (a) and (b) for $RWP_{20K}$ and $VN_{2K}$ datasets, respectively. Based on our results, ReachGrid approach is comparable with ReachGraph for the cases in which the query interval is small. The reason is that with such cases, a small portion of contact network should be traversed which is placed on consecutive blocks on disk and can efficiently retrieved from disk by ReachGrid. Another important observation is that in addition to the query interval size, the distribution of objects also affects the performance of ReachGrid. With $VN_{2k}$ dataset, the objects are located on road network and within the small portion of entire environment $E$ as opposed to $RWP_{20k}$ dataset for which the objects are almost uniformly distributed in $E$. As the result, with $VN_{2k}$ dataset ReachGrapth approach significantly outperforms ReachGrid (on average 63%). The reason is that ReachGrid spatial grid cannot leverage spatial locality for non-uniform objects distributions.

We also compare the CPU time of both approaches which is the time it takes by the algorithms while ignoring retrievals from disk. The result is shown in Figure 15 for $RWP_{20k}$ and $VN_{2K}$ datasets. As expected, ReachGraph has significantly lower CPU time because of extensive offline precalculations and hence avoiding spatiotemporal joins at the query time.

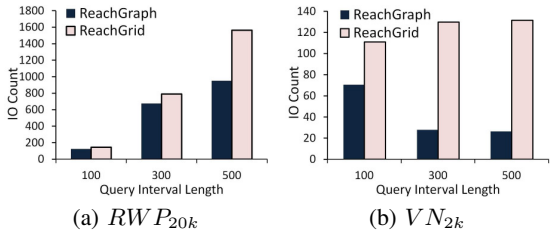

**Figure 14: ReachGrid vs. ReachGraph**

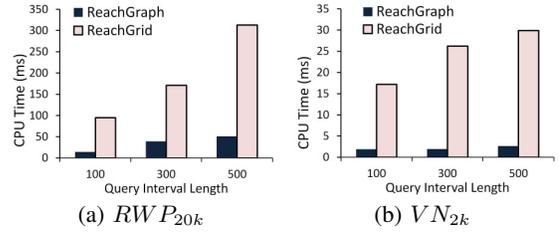

**Figure 15: CPU time**

| (a) Memory Resident Contact Datasets | | | (b) Disk Resident Contact Datasets | | |
|---|---|---|---|---|---|
| Dataset | GRAIL (runtime) | RG (runtime) | Dataset | GRAIL (IO Count) | RG (IO Count) |
| $VN_{2k}$ | 3.5ms | 9.0ms | $VN_{2k}$ | 213 | 49 |
| $RWP_{20k}$ | 60ms | 39ms | $RWP_{20k}$ | 6790 | 570 |

**Table 5: GRAIL vs. ReachGraph (denoted by RG)**

## 6.4 Comparison with Graph Reachability

Here, we compare ReachGraph query processing with the existing graph reachability techniques. In particular, we compare our approach with GRAIL [18]. First, we consider contact datasets which reside in memory. We compare the performance of ReachGraph and GRAIL on $RWP_{20K}$ and $VN_{2K}$ contact datasets with $|T|$=1000, which are memory resident datasets. GRAIL takes $D_N$ as input and verifies whether the query source is reachable to the query destination. Table 5 (a) shows the results of this comparison in terms of runtime for random queries with the interval length of 300. GRAIL converges to simple DFS for reachability queries when source and destination are reachable. Therefore, our approach outperforms GRAIL for $VN_{2K}$ while this is not the case for $RWP_{20K}$ because of the existence of more pairs of reachable objects in $VN_{2K}$ than $RWP_{20K}$. With $RWP_{20K}$, GRAIL is 30% faster than ReachGraph. In sum, we conclude that our approach is comparable with GRAIL for memory resident contact datasets.

Next, we adopt GRAIL for disk-resident contact datasets and subsequently compare the performance of GRAIL and ReachGraph in terms of number of IOs for disk-resident contact networks. To this end, we issue the same queries but on the disk resident contact datasets. We assume that with GRAIL the vertices are placed on disk in the same order they are generated during contact network construction. The results are shown in Table 5 (b). As expected, our approach significantly outperforms GRAIL for disk-resident datasets. In particular, it outperforms GRAIL for 76% and 88% for $VN_{2K}$ and $RWP_{20K}$ datasets, respectively.

## 7. DISCUSSION

In this section, we briefly discuss how our algorithms can be potentially extended to address more generic contact-network reachability problems where the definition of "contact" is partly different. In particular, we discuss two cases. First, we consider *uncertain* contact networks and then, we focus on *non-immediate* contacts.

Two objects $o_i, o_j \in O$ make an uncertain contact with probability $p$, when their distance is less than $d_T$ and transmit an item with the probability of $p$. For example, with most viral diseases an individual can infect another one with a disease with some probability of $p$ once in proximity, where $p$ depends on various factors such as the distance between the individuals. Accordingly, a contact path $P$ is also probabilistic with a probability which is the multiplication of the probability of contacts in $P$. We say, $o_j$ is reachable from $o_i$ during $T_p$ if a contact path exists from $o_i$ to $o_j$ with the probability of at least $p_T$. We term the ReachGraph for uncertain contact networks U-ReachGraph and briefly explain how one can extend ReachGraph to U-ReachGraph. We skip the details on the ReachGrid extension due to lack of space.

Both index construction and query processing with U-ReachGraph are different from those of the ReachGraph. For index construction, each edge $e$ of the TEN model is associated with a weight representing the probability of contact between the objects represented by $e$ endpoints. For reduction, although the first step cannot be applied to the TEN model for uncertain networks (unless all the edges within a connected component are associated with the probability of one), the second reduction step readily applies to uncertain contact networks assuming that the contact probabilities are taken into consideration. For augmentation, a long edge $e$ from $v_i$ to $v_j$ where $v_i, v_j \in D_N$ must also be associated with a probability $p$, where $p$ is the probability of the contact path with the highest probability from $v_i$ to $v_j$ during a time interval spanned by $e$. With ReachGraph, we adopted graph traversal approaches such as BFS to process a reachability query, whereas with U-ReachGraph the contact path probability is also important. Accordingly, with U-ReachGraph we adopt graph shortest path algorithms to verify whether a contact path with the probability of at least $p_T$ exists from query source to destination during query interval.

A non-immediate contact between $o_i$ and $o_j$ occurs when the distance between the location of $o_i$ at time $t$ and that of $o_j$ at time $t'(t \leq t')$ is less than a threshold $d_T$ and $|t' - t| \leq T_t$. $T_t$ is the lifetime of the item initiated by objects and $[t, t']$ is the contact validity interval. For example, a person $u$ carrying a virus may spread it in a bus at time $t$ and get off the bus. Later on, another person $v$ may get on the same bus at time $t'$ and become infected. The definitions in Section 3 readily apply to non-immediate contact networks. Moreover, ReachGrid and ReachGraph can also be readily adopted for non-immediate contact network with one exception. With regular contact networks, we perform spatiotemporal join between object trajectories to extract contacts between objects, whereas with non-immediate contact network the replicated trajectories should be joined to produce contacts between objects.

## 8. CONCLUSION AND FUTURE WORK

In this paper, for the first time we introduced and studied the problem of reachability query in disk-resident spatiotemporal contact networks. We proposed two different indexing approaches, ReachGrid and ReachGraph, to enable efficient reachability query processing. We have conducted an empirical study with both real and synthetic datasets to evaluate our proposed techniques. The experimental results show that our proposed techniques outperform the existing reachability query processing approaches in contact networks by 76% on average.

In the future, we plan to continue our study on reachability in uncertain and non-immediate contact networks. Moreover, we intend to extend the techniques proposed in this paper to consider item transmission delay in contacts. Finally, we plan to extend our proposed approaches to be applicable in cloud-computing environments to further enhance the efficiency of query processing.

## 9. ACKNOWLEDGMENTS

This research is supported in part by Award No. 2011-IJ-CX-K054 from National Institute of Justice, Office of Justice Programs, U.S. Department of Justice, as well as by the NSF grant IS-1115153, the USC Integrated Media Systems Center (IMSC), and unrestricted cash and/or equipment gifts from Northrop Grumman, Microsoft and Qualcomm. The opinions, findings, and conclusions or recommendations expressed in this publication are those of the authors and do not necessarily reflect those of the Department of Justice and the National Science Foundation. We also thank Professor Lin Zhang at Tsinghua university for providing us with the real dataset of vehicles movement in the city of Beijing.